\documentclass[letterpaper,twocolumn,english,amssymb,aps,prl,floats]{revtex4}
\usepackage[T1]{fontenc}
\usepackage[latin1]{inputenc}
\usepackage{graphicx}
\usepackage{amssymb}

\makeatletter

\usepackage{babel}
\makeatother
\begin{document}

\title{Dissipative dynamics of spins in quantum dots}

\author{Harry Westfahl Jr.$^{1}$, Amir O. Caldeira$^{2}$, Gilberto Medeiros-Ribeiro$^{1}$,
and Maya Cerro$^{1,2}$ }

\affiliation{$^{1}$Laboratório Nacional de Luz Síncrotron - ABTLuS, Caixa Postal
6192, Campinas, SP 13043-090, Brazil}

\affiliation{$^{2}$Instituto de Física ''Gleb Wataghin\char`\"{}, Universidade
Estadual de Campinas, Caixa Postal 6165 , Campinas, SP 13083-970,
Brazil }

\date{\today{}}


\begin{abstract}
We present a theory for the dissipation of electronic spins trapped
in quantum dots due to their coupling to the host lattice acoustic
phonon modes. Based on the theory of dissipative two level systems
for the spin dynamics, we derive a relation between the spin dissipative
bath, the electron confinement, and the electron-phonon interaction.
We find that there is an energy scale, typically smaller than the
electronic lateral confinement energy, which sets the boundary between
different dissipative regimes .
\end{abstract}
\maketitle
Manipulating quantum states of matter to achieve new ways of processing
information has been a subject of intense research in the last decade
\cite{Chuang00}. There is now a wide variety of proposals to take
advantage of the internal transformations of quantum systems to perform
the so-called quantum information processing (QIP) \cite{Loss98}.
In particular, condensed matter systems such as SQUIDS and semiconductor
quantum dots are seen as possible candidates for future implementations
of QIP devices. In these systems, it is possible to find or create
well defined two-level quantum states that potentially fulfill the
minima criteria proposed by Di Vicenzo to be quantum bits (qubits)
\cite{Vicenzo95}. However, apart from the technical difficulties
to implement acceptable qubits, nature imposes another limitation
which is ubiquitous. This comes from the impossibility of perfectly
isolating a complex quantum system from its environment that results
in the loss of quantum coherence \cite{amir83,Leggett87}. 

In this paper we investigate the decoherence of electronic spins in
quantum dots due to their coupling to the acoustic phonon modes of
the host lattice. In few-electron quantum dots the electronic spin
is a good candidate for a qubit and the confinement of the electronic
wave function plays a major role in isolating the spins from most
energy relaxation channels. In fact, using a second order perturbation
approach, Khaetskii and Nazarov \cite{Khaetskii} have evaluated the
spin-flip transition rate between Zeeman sublevels due to its coupling
to acoustic phonons. They showed that, since angular momentum conservation
requires the spin-flip process to involve a virtual transition between
excited orbital states, the spin-flip rates are suppressed as $\omega_{0}^{-4}$,
where $\omega_{0}$ is the lateral confinement energy. According to
their result, even for $\omega_{0}\sim1\, meV$ , the spin-flip rates
can be of the order of milliseconds.

Quantum dots can be fabricated by confining a two dimensional electron
gas of a semiconductor heterostructure into a region of the order
of the Fermi wavelength. In the {}``lateral quantum dots'' (LQD)
this confinement is done laterally, by means of surface electrostatic
metallic gates. In {}``vertical quantum dots'' (VQD) the electron
gas is confined vertically by etching techniques that create a circular
pillar heterostructure. The typical lateral confinement length achieved
with these fabrication techniques are of the order of hundreds of
nanometers, which leads to $\omega_{0}\approx1\, meV$ for LQD \cite{Hanson03}
and $\omega_{0}\approx3-5\, meV$ for the VQD \cite{Fujisawa02}.
Much higher confinement energies can be achieved in self-assembled
quantum dots (SAQD) which consist of ensembles of dislocation free
semiconductor nano-crystals embedded into a semiconducting matrix
of different band gap \cite{BimbergBook}. For InAs:GaAs SAQD \cite{Gilberto99}
the electronic wave function extent is of the order of $50\,\textrm{Å}$,
resulting in $\omega_{0}\approx50\, meV$. For all the above systems
single electron charging is easily observable due to their large Coulomb
blockade gap, which can be of the order of $2\, meV$ for LQD and
VQD and as large as $20\, meV$ for SAQD.

Although no experiment has so far been able to determine directly
the decoherence rate, $\frac{1}{T_{2}}$, of spins in quantum dots,
there are several attempts to indirectly determine lower bounds to
the spin relaxation rate $\frac{1}{T_{1}}$. Paillard and co-workers
\cite{Paillard01} have studied the time-resolved photoluminescence
of InAs:GaAs SAQD and observed that, within the time scale of an exciton
lifetime, the carrier spins are totally frozen. Fujisawa and co-workers
\cite{Fujisawa02} have demonstrated that orbital sublevel transitions
of VQD involving spin-flips have relaxation times that are 4 to 5
orders of magnitude longer than those which do not involve spin-flips.
This gives an idea of the degree of isolation of the spins in the
VQD. More recently, Hanson and co-workers \cite{Hanson03} used short
voltage pulse sequences to measure the relaxation time of spins in
LQD and concluded that the lower bound of $T_{1}$ is $50$$\mu s$
for a magnetic field of $7.5\, T$. 

In quantum dots, the most important channels of dissipation for the
spin are indirect and via the spin-orbit interaction. The orbital
dissipative dynamics on its turn is dominated by the electron-phonon
interactions. It is thus desirable to have a theoretical description
which can connect the spin and orbital dissipation channels. A similar
problem of indirect dissipation appears on a completely different
context of electron tunneling between the atoms of a diatomic molecule
embedded in a viscous environment \cite{Onuchic95}. Here we analyze
the spin dissipation in a quantum dot from the same perspective. We
derive an effective {}``bath'' spectral density seen by the spins
that results from the spin-orbit coupling and the orbital damping.

Following previous theoretical works we treat the electrons in the
effective mass approximation and consider the confining potential
for the envelope wave function as parabolic. This was shown \cite{Matagne02}
to be a good approximation for the LQD and VQD with low electronic
filling. Also, for SAQD this phenomenological model describes the
orbital electronic density of states probed by magneto-capacitance
measurements with fine accuracy up to the 3th excited level of the
dot \cite{Gilberto99}. We further assume that the harmonic frequency
in the direction perpendicular to the quantum dot plane, $\omega_{\perp}$,
is much higher than the lateral harmonic frequency $\omega_{0}$.
Therefore the relevant low energy orbital degrees of freedom are in
the $x-y$ plane whereas the orbital dynamics in the $z$ direction
is practically frozen. This is a reasonable assumption even for typical
InAs:GaAs SAQD since $\frac{\omega_{\perp}}{\omega_{0}}\approx8-10$
\cite{Gilberto99}. For the spin degrees of freedom, besides the Zeeman
term arising from an external magnetic field in the $z$ direction,
we include the Dresselhaus spin-orbit interaction projected on the
$x,y$ plane \cite{Ivchenko95} which is responsible for the coupling
between the spins and the dissipative phonon bath. Thus, apart from
a zero point energy on the $z$ direction, the spin-orbit Hamiltonian
will be\begin{eqnarray}
H_{SO} & = & -\frac{\Delta}{2}\sigma_{z}+\omega_{0}\left(a_{x}^{\dagger}a_{x}+\frac{1}{2}\right)-\beta\sigma_{x}P_{x}\nonumber \\
 &  & +\omega_{0}\left(a_{y}^{\dagger}a_{y}+\frac{1}{2}\right)+\beta\sigma_{y}P_{y}\label{eq:a}\end{eqnarray}
where $\Delta=g\mu_{B}B_{z}$ (we use $\hbar=1$), $\beta\equiv\gamma_{c}\left\langle k_{z}^{2}\right\rangle =\gamma_{c}m^{*}\omega_{\perp}$
(with $\gamma_{c}$ being the Kane parameter \cite{Ivchenko95}),
$m^{*}$ is the electron effective mass and $g$ its gyromagnetic
factor. The operators $a_{x\left(y\right)}$ are the usual ladder
operator for the $x\left(y\right)$ direction. Notice that although
we set an external field for the spins, we are neglecting any diamagnetic
contribution for the orbital degrees of freedom like in the Fock-Darwin
description. This simplification allows us to separate the degrees
of freedom in the $x$ and $y$ directions and can be well justified
for $\omega_{0}\gg\frac{\omega_{c}}{2}$, where $\omega_{c}=eB/\left(m^{*}c\right)$.
One should also notice that, as opposed to the notation on reference
\cite{Leggett87}, $\Delta$ here plays the role of a {}``tunneling''
field rather than the {}``bias'' field, even though it refers to
the $z$ direction. This happens because dissipation occurs only in
the $x-y$ plane of the dot.

The electron-phonon coupling in this restricted subspace can be written
as\begin{equation}
H_{e-ph}=\sum_{\mathbf{q},\lambda}\omega_{\mathbf{q},\lambda}b_{\mathbf{q},\lambda}^{\dagger}b_{\mathbf{q},\lambda}+\frac{C_{\mathbf{q},\lambda}}{\sqrt{V}}e^{i\mathbf{q}.\mathbf{r}}\left(b_{\mathbf{q},\lambda}^{\dagger}+b_{\mathbf{q},\lambda}\right)\label{eq:b}\end{equation}
where $C_{\mathbf{q},\lambda}$ is the electron-phonon coupling for
phonons with polarization $\lambda$ and frequency $\omega_{\mathbf{q},\lambda}$
, and $\mathbf{r}=\left(x,y,z\right)$ is the electron position operator.
Here we consider only the piezoelectric and deformation potential
interactions with acoustic phonon modes in zinc-blende structures
\cite{Cardona}. It can be shown that within the linear response approximation
for the phonon system the electron-phonon Hamiltonian (Eq.\ref{eq:b})
is mapped into the \emph{bath of oscillators} model \cite{amir83}
with the spectral function given by \[
J_{s}\left(\omega\right)=m^{*}\omega_{D}^{2}\delta_{s}\left(\frac{\omega}{\omega_{D}}\right)^{s}\theta\left(\omega_{D}-\omega\right)\,,\]
where $s=3$ for the piezoelectric interaction, with dimensionless
coupling $\delta_{3}=\frac{\left(e_{m}\right)_{14}^{2}\omega_{D}}{35\pi m^{*}\rho}\left(\frac{4}{3v_{t}^{5}}+\frac{1}{v_{l}^{5}}\right)$,
and $s=5$ for the deformation potential, with $\delta_{5}=\frac{a_{c\Gamma}^{2}\omega_{D}^{3}}{2\pi\rho m^{*}v_{l}^{7}}$,
where $\omega_{D}$ is the Debye frequency, $v_{l}$ and $v_{t}$
are the longitudinal and transverse sound velocities respectively,
$\rho$ is the material density, $\left(e_{m}\right)_{14}$ is the
electromechanical tensor for zinc-blende structures \cite{Cardona},
and $a_{c,\Gamma}$ is the deformation potential in the $\Gamma$
point \cite{Cardona}. $\theta$ is the Heaviside step function.

Now, since the spin degree of freedom is coupled to the orbital motion
of the electron, we can adopt the prescription of reference \cite{Onuchic95}
to extract the spectral function of the effective heat bath to which
the spin is now coupled. As opposed to the case of reference \cite{Onuchic95}
where both the {}``spin'' and the phonon degrees of freedom are
coupled to the position of the orbit, here we have the phonon coupled
to the position (Eq.\ref{eq:b}) and the spin to the momentum of the
electron (Eq.\ref{eq:a}). This leads to a significant change on the
{}``effective bath'' spectral function, namely \begin{equation}
J_{eff}\left(\omega\right)=m^{*}\beta^{2}\frac{\delta_{s}\left(\frac{\omega}{\omega_{D}}\right)^{s+2}}{Z\left(\omega\right)^{2}+\delta_{s}^{2}\left(\frac{\omega}{\omega_{D}}\right)^{2s}}\theta\left(\omega_{D}-\omega\right),\label{eq:phi}\end{equation}
where $Z\left(\omega\right)\equiv\left(\frac{\omega_{0}}{\omega_{D}}\right)^{2}-\left(\frac{\omega}{\omega_{D}}\right)^{2}\left(1+\delta_{s}\,\phi_{s}\left(\frac{\omega}{\omega_{D}}\right)\right)$,
and $\phi_{s}\left(x\right)\equiv\frac{2}{\pi}\mathcal{P}\int_{0}^{1}x^{s}/\left(y^{3}-yx^{2}\right)dy=-\frac{x^{s-2}}{\pi}\left(B\left(x,s,0\right)+\left(-1\right)^{s}B\left(-x,s,0\right)\right)$,
with $B$ being the generalized incomplete \emph{beta function}. For
$x\ll1$ we can approximate $\phi_{s}\left(x\right)\simeq\frac{2}{\pi}\left(\frac{1}{s-2}+\frac{x^{2}}{s-4}\right)$

For most semiconductors used in the fabrication of quantum dots the
dimensionless constants $\delta_{3}$ and $\delta_{5}$ are of the
order of $10^{2}$ and $10^{6}$ respectively. For instance, using
the bulk physical paramenters \cite{Cardona} we obtain, in $GaAs$
$\delta_{3}=355$ and $\delta_{5}=1.95\times10^{6}$ and in $InAs$
$\delta_{3}=149$ and $\delta_{5}=5.03\times10^{6}$. Furthermore,
the typical frequency for the spin dynamics is much smaller than $\omega_{D}\approx30-50\, meV$,
which suggests an asymptotic analysis of $J_{eff}\left(\omega\right)$.
In order to do that we should first notice that $J_{eff}\left(\omega\right)$,
Eq. \ref{eq:phi}, is peaked at $\omega=\Omega_{s}$, where $\Omega_{s}$
is defined as the solution of $Z\left(\Omega_{s}\right)=0$. For $\omega_{0}\lesssim\omega_{D}$
this solution is given by: \[
\Omega_{s}\approx\omega_{0}\sqrt{\frac{(s-2)}{(s-2)+\frac{2\delta_{s}}{\pi}}}\,.\]

\begin{figure}
\begin{center}\includegraphics{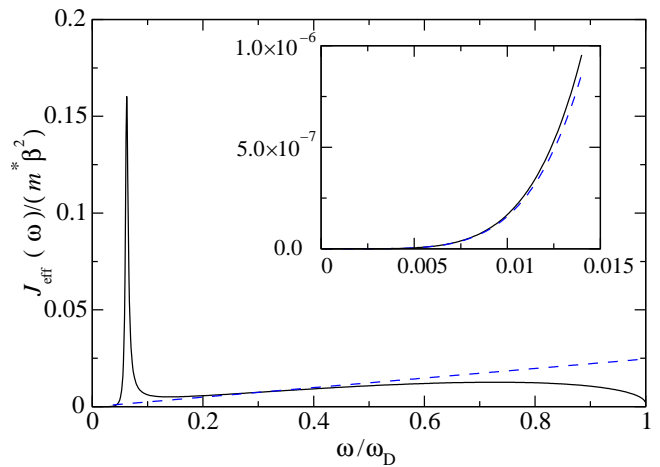}\end{center}

\caption{\label{cap:Behavior} Full lines, exact $J_{eff}\left(\omega\right)$
($s=3$, $\frac{\omega_{0}}{\omega_{D}}=0.5$ and $\delta_{3}=100$),
dashed lines, asymptotic behavior according to Eq. \ref{eq:HF} and
Eq. \ref{eq:LF} (inset). }
\end{figure}

In the weak coupling limit, i.e., $\delta_{s}\ll1$, $\Omega_{s}\sim\omega_{0}\left(1-\frac{\delta_{s}}{\pi\left(s-2\right)}\right)$
and the resonance at $\Omega_{s}$ corresponds to the harmonic frequency
of the orbital dynamics, shifted by the electron-phonon coupling.
However, for $\delta_{s}\gg1$, the peak originally centered at the
lateral confinement frequency, $\omega_{0}$, is drastically shifted
to $\Omega_{s}=\omega_{0}\sqrt{\frac{(s-2)\pi}{2\delta_{s}}}$. In
the low frequency range, defined by $\omega\ll\Omega_{s}$ and $\frac{\omega}{\omega_{D}}\ll\left(\frac{\omega_{0}}{\omega_{D}}\frac{1}{\delta_{s}}\right)^{1/s}$,
we obtain\begin{equation}
J_{eff}\left(\omega\right)\approx m^{*}\beta^{2}\delta_{s}\left(\frac{\omega_{D}}{\omega_{0}}\right)^{4}\left(\frac{\omega}{\omega_{D}}\right)^{s+2}\,.\label{eq:LF}\end{equation}
Note that in this limit the effective spectral function felt by the
spins is always super-ohmic, with a power $s+2$. On the other hand,
in the high frequency limit, $\Omega_{s}\ll\omega\ll\omega_{D}$,
the spectral function can be approximated by\begin{equation}
J_{eff}\left(\omega\right)\approx m^{*}\beta^{2}\frac{\pi^{2}\left(s-2\right)^{2}}{4\delta_{s}}\left(\frac{\omega}{\omega_{D}}\right)^{s-2}\,.\label{eq:HF}\end{equation}
In this limit the spectral function has a power $s-2$ that can even
be sub-ohmic. The behavior of $J_{eff}\left(\omega\right)$ is sketched
in figure \ref{cap:Behavior} for the piezoelectric interaction ($s=3$). 

All this complex structure of the effective spectral function will
provide us with a new physics for the system. Within the adiabatic
renormalization scheme, the fast modes of the bath, i.e. those with
frequency much higher than $\Delta$, can quickly adjust to the spin
flip motion and are successively integrated out by a Born-Oppenheimer
approximation \cite{Leggett87}.

At long times the spin flips coherently with a renormalized Zeeman
frequency $\Delta_{r}$, dressed by a Franck-Condon (FC) factor, which
represents the polarization cloud of the high frequency phonon-orbit
complex, given by :\begin{equation}
\Delta_{r}=\Delta e^{-F_{s}\left(\Delta_{r}\right)}\,,\label{eq:Dr}\end{equation}
where $F\left(z\right)=\frac{1}{2}\int_{pz}^{\infty}d\omega\frac{J_{eff}\left(\omega\right)}{\omega^{2}}\coth\left(\frac{\omega}{2T}\right)$
is the FC factor \cite{Leggett87} with $p$ being an unimportant
dimensionless number much larger than 1. In the low frequency limit
$z\ll\Omega_{s}$ the FC factor is dominated by the region $\omega\sim\Omega_{s}$
where $\frac{J_{eff}\left(\omega\right)}{\omega^{2}}$ can be well
approximated by a Lorentzian of weight $\frac{1}{\left|Z^{\prime}\left(\Omega_{s}\right)\right|}$
and width $\epsilon_{s}=\frac{\omega_{D}\delta_{s}\Omega_{s}^{s}}{\left|Z^{\prime}\left(\Omega_{s}\right)\right|}$.
For $\delta_{s}\gg1$, to the leading order in $\delta_{s}^{-1}$
we have $Z'\left(\Omega_{s}\right)=2\frac{\omega_{D}}{\Omega_{s}}\left(\frac{\omega_{0}}{\omega_{D}}\right)^{2}$
and thus $\epsilon_{s}\simeq\frac{\omega_{D}\pi\left(s-2\right)}{4}\left(\frac{\omega_{0}}{\omega_{D}}\sqrt{\frac{(s-2)\pi}{2\delta_{s}}}\right)^{s-1}$.
Since $\Delta_{r}\leq\Delta$, the solution for $\Delta\ll\Omega_{s}$
involves integration over the entire spectrum of $J_{eff}\left(\omega\right)$,
where most of the weight is concentrated around the peak centered
at $\Omega_{s}$, and thus the lower limit of integration can be extended
to zero, yielding\[
\Delta_{r}\approx\Delta\exp\left\{ -\frac{\Omega_{s}}{2T_{0}}\coth\left(\frac{\Omega_{s}}{2T}\right)\right\} \,,\]
where $T_{0}=\frac{2\omega_{0}^{2}}{\pi m^{*}\beta^{2}}$. For temperatures
$T\gg\Omega_{s}$ we then have $\Delta_{r}\sim\Delta\exp\left\{ -\frac{T}{T_{0}}\right\} $.
From this point onwards we restrict our discussion to the piezoelectric
electron-phonon coupling which leads to a larger bath spectral density
at low frequencies. In the low frequency limit, defined by $\Delta\ll\Omega_{3}$
, the dynamics is dominated by a super-ohmic relaxation with a power
$s+2$ (see Eq. \ref{eq:LF}). This allows the spin to present coherent
damped oscillations with a decoherence rate \cite{Leggett87} $\frac{1}{T_{2}}=2J_{eff}\left(\Delta_{r}\right)\coth\left(\frac{\Delta_{r}}{2T}\right)$,
given by\[
\frac{1}{T_{2}}\sim2m^{*}\beta^{2}\delta_{3}\left(\frac{\omega_{D}}{\omega_{0}}\right)^{4}\left(\frac{\Delta_{r}}{\omega_{D}}\right)^{5}\coth\left(\frac{\Delta_{r}}{2T}\right)\,.\]
Except for the renormalized Zeeman frequency $\Delta_{r}$, this is
essentially the spin-flip rate from Khaetskii and Nazarov \cite{Khaetskii}
which, for $GaAs$ quantum dots with $m^{*}\beta^{2}\sim O\left(\mu eV\right)\sim10^{9}s^{-1}$,
$\omega_{0}=1\, meV$ and $\Delta=0.025\, meV$, gives decoherence
times of the order of milliseconds . One should note however that,
since $\Omega_{3}\sim O\left(0.1\,\omega_{0}\right)$, for Zeeman
splittings in the range $\Delta\sim0.1-1\, meV$ the perturbative
result is only applicable for dots with lateral confinement energies
$\omega_{0}\gg1\, meV$. This is usually the case then for SAQD, in
which one can have $\omega_{0}\simeq50\, meV$. For the LQD and VQD
$\omega_{0}\sim O\left(1\, meV\right)$ and thus $\Delta$ can go
beyond this perturbative limit. For $\Delta\sim\Omega_{3}$, the relaxation
is dominated by the resonance of the spin with the orbit-plus-phonons
composite at frequency $\Omega_{3}$. This resonance has a linewidth
of $\epsilon_{3}\simeq\left(\frac{\omega_{0}^{2}}{\omega_{D}}\right)\frac{\pi^{2}}{8\delta_{s}}$
which qualitatively determines the decoherence rate of the spin in
this intermediate range of frequencies.

In the higher frequency range $\Omega_{s}\ll\Delta\ll\omega_{D}$
the integral on equation (\ref{eq:Dr}) is dominated by the high frequency
part of $J_{eff}\left(\omega\right)$. For $s>3$ this integral is
algebraic and only leads to a small correction. However, if $s\leq3$
the behavior is qualitatively different and dominated by the lower
limit of the integration which has an {}``infrared divergence''.
For the piezoelectric interaction ($s=3$), it behaves at low temperatures
$T\ll\Theta_{D}$ as\[
\frac{\Delta_{r}}{\omega_{D}}=\left(\frac{\Delta}{\omega_{D}}\right)^{\frac{1}{1-K}}\,,\]
where $K=\frac{m^{*}\beta^{2}}{\omega_{D}}\frac{\pi^{2}\left(s-2\right)^{2}}{8\delta_{s}}\ll1\,.$
In this {}``high frequency region'', $\Omega_{s}\ll\Delta\ll\omega_{D},$
that can be achieved for $\omega_{0}\ll1\, meV$ ({}``large'' quantum
dots), the decoherence will be dominated by an ohmic-like dissipation
given by Eq. \ref{eq:HF}, which yields a decoherence rate\[
\frac{1}{T_{2}}\sim m^{*}\beta^{2}\frac{\pi^{2}}{2\delta_{s}}\left(\frac{\Delta_{r}}{\omega_{D}}\right)\,.\]

\begin{figure}
\begin{center}\includegraphics{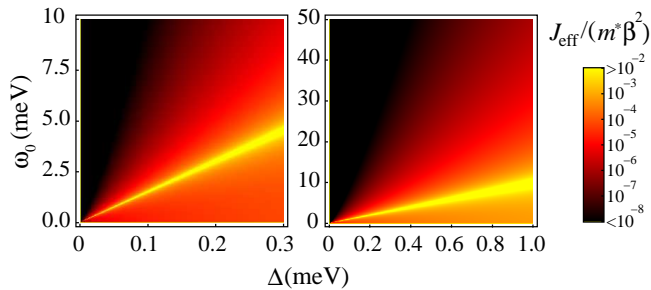}\end{center}

\caption{\label{cap:Grayscale}Density map of $J_{eff}/\left(m^{*}\beta^{2}\right)$
from piezoelectric interaction ($s=3$) for $GaAs$ (left) and $InAs$
(right) QD. }
\end{figure}

Figure \ref{cap:Grayscale} shows the behavior of $J_{eff}$ for $GaAs$
and $InAs$ quantum dots. The darkest region to the left (right) of
the brighter stripe (corresponding to the peak) is the super-ohmic
(ohmic) dissipation. A perpendicular magnetic field will increase
the lateral confinement according to the Fock-Darwin energy $\omega_{FD}=\sqrt{\omega_{0}^{2}+\frac{1}{4}\omega_{c}^{2}}$.
This is the case for the experiment on reference \cite{Hanson03}
where the magnetic fields are of the order of $10\, T$ and thus the
cyclotron frequency $\omega_{c}$ contribution can be higher than
$\omega_{0}$. The density map of Figure \ref{cap:Grayscale} is then
a guide for one to reach lowest decoherence rates given the lateral
confinement energy of the dot and the Zeeman splitting $\Delta$.

Thus, we conclude that the piezoelectric electron-phonon coupling
only leads to large decoherence times, as predicted by the second-order
perturbation theory, if the $\frac{\Delta}{\omega_{0}}\ll\frac{1}{\sqrt{\delta_{3}}}$.
Otherwise the decoherence rate has a completely different behavior
leading to much higher decoherence rates. This suggests that as far
as the decoherence due to acoustic phonons is concerned, SAQD better
decouple the electronic spin degree of freedom from the environment.
Nevertheless one should note that there are other decoherence mechanisms
due to phonons. In reference \cite{woods02} other electron phonon
mechanisms were considered extending the analysis of Khaetskii and
Nazarov \cite{Khaetskii}. It has also been argued \cite{Li98,Verzelen00,Sauvage02}
that in SAQD, a resonance between the lateral confinement frequency
and the longitudinal optical (LO) phonon mode can lead to a more efficient
channel of orbital energy dissipation. In this scenario the orbital
relaxation rate should be determined by the LO relaxation rate $\Gamma_{op}$(in
bulk $GaAs$ $\Gamma_{op}^{-1}\approx7\, ps$). Since the approach
developed here connects the spin dissipation to the orbital dissipation
and thus must also apply to these other orbital dissipation mechanisms
if we replace $J\left(\omega\right)$ by the appropriate bath spectral
function. 

This research was partially supported by Hewlett-Packard Brazil. The
authors also acknowledge partial support from Conselho Nacional de
Desenvolvimento Científico e Tecnológico (CNPq) and Fundação de Amparo
à Pesquisa no Estado de São Paulo (FAPESP).

\end{document}